# Gamma Background Measurements at BEO Moussala

A. Mishev* and E. Hristova

Institute for Nuclear Research and Nuclear Energy, Bulgarian Academy of Sciences

72 Tsarigradsko chaussee, Sofia 1284, Bulgaria

**Abstract:**

*Recent measurements of radiation gamma background at high mountain altitude, namely at Basic Environmental Observatory Moussala (42°10' 45" N; 23° 35'07"E, 2925m a.s.l.) are carried out. The measurements are fulfilled with several different devices, namely IGS-421 gamma probe and MDU Liulin spectrometer. The used instrumentation with their characteristics is described. The obtained results are compared with previously obtained measurements carried out with NaI SAPHYMO gamma probe. In addition the measurements are compared with TLD data. The obtained results from measurements are widely discussed. The contribution of galactic cosmic rays to dose rate is estimated on the basis of a numerical model. The model is based on a full Monte Carlo simulation of cosmic ray induced nuclear-electromagnetic-muon cascade in the atmosphere. The simulation is carried out with CORSIKA 6.52 code using FLUKA 2006 and QGSJET II hadron interaction models. The application of the model is briefly discussed.*

## 1. Introduction

The radioactive elements and their radiation are indispensable part of the nature. Their influence on live organisms is imminent and very important to study. The natural radioactivity is due mainly to cosmic radiation and the availability of natural radio-nuclides (terrestrial sources) in the environment [1]. A significant part of the exposure is due to sources in food and water, which are incorporated in the body, as well to sources in building materials and other products that contain radioactive elements. Another significant contribution comes from the radon gas and its radioactive progeny. It is released from the Earth's crust and subsequently decays into radioactive atoms that become attached to airborne dust, aerosols and other particulates. Another contribution arises from the radioactive atoms produced in the bombardment of atoms in the upper atmosphere by high-energy cosmic rays, the so called cosmogenic radionuclides. Summarizing the natural radioactivity can be placed in three general categories:

* *Corresponding author:*

*Dr. Alexander Mishev, Institute for Nuclear Research and Nuclear Energy, Bulgarian Academy of Sciences Tel: ++ 359 2 9746310; Fax: ++ 359 2 9753619; e-mail: mishev@inrne.bas.bg*

1. Primordial - been since the creation of the Earth
2. Cosmogenic - formed as a result of cosmic ray interactions
3. Human produced - enhanced or formed due to human activity

About 3 % of the background radiation comes from other man-made sources. The level of natural background radiation varies depending on location over the world. In some areas the level is significantly higher than the average [2, 3].

During the years the high mountain observatories have been exploited for environmental studies and monitoring. The advantages of such places are the small anthropogenic influence, which permits to monitor trans-border transport and pollution. The Basic Environmental Observatory (BEO) Moussala (42°10' 45" N; 23° 35'07"E) is located on the top of the highest mountain on the Balkan Peninsula, namely at 2925m above sea level. This is a privileged place for such type of investigations. This is one of the most proper places in the region of Balkans. It is with small anthropogenic influence. It gives an excellent possibility for environmental monitoring [4].

The general aim of BEO Moussala is complex aerospace and environmental monitoring. The existing devices at BEO Moussala permit to study changes and processes in the atmosphere, climate parameters, space weather [5], atmospheric transparency [6], cosmic ray [7] etc... A significant part of BEO Moussala monitoring is related to radiation gamma background measurements [8]. At present the radiation gamma background measurements are fulfilled with several devices (Fig.1), namely IGS-421 gamma probe and Liulin spectrometer. At the same picture the previously used for radiation gamma background measurements device – SAPHYMO gamma probe is also shown.

## 2. Instrumentation

After the nuclear reactor accident in Chernobyl in 1986, the majority of the countries of the European Union established monitoring networks measuring outdoor dose rates. The general aim is to provide a system for early warning. Basically the measured data are composite values: gamma-dose rate due to terrestrial, cosmic and artificial radiation sources. In most cases the data include some instrument background. At the same time the data, stored in the database potentially contain valuable information about spatio-temporal variations, term which can be used for the validation of atmospheric transport models and other atmospheric tracer applications. Several different devices for radiation gamma background measurements are operational at BEO Moussala. The aim is to provide continuous recordings of radiation gamma background. Taking into account the specific climate conditions at high mountain, specifically during the winter period, duplication of these measurements is necessary.

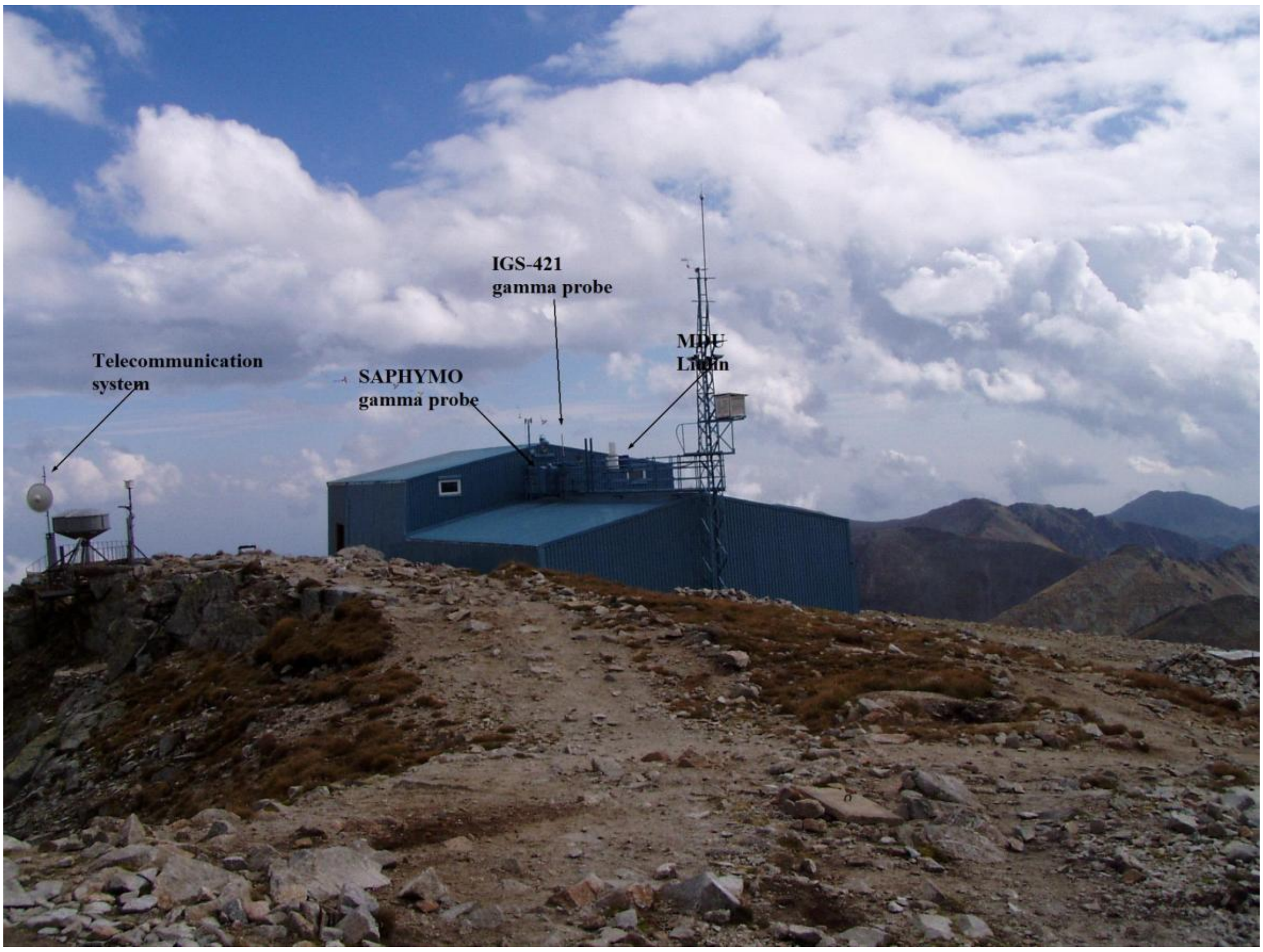

Fig. 1 Basic Environmental Observatory (BEO) Moussala

### 2.1 SAPHYMO gamma probe

A well known method for gamma ray detection involves the use of crystal scintillators such as NaI(Tl). When it is used as a gamma-ray detector, the scintillator does not directly detect the gamma-rays. The gamma-rays produce charged particles in the scintillator crystals, which interact with the crystal and emit photons. These low energy photons are subsequently collected by photomultiplier. Historically the first device installed at BEO Moussala was SBN-90 SAPHYMO gamma probe based on NaI(Tl) detector (Fig.2) supplied by French electrical company (EDF). In fact a network of four devices distributed mainly in South of Bulgaria was established. The energy range of the probe is between 50 keV and 3 MeV and gives the possibility to measure dose rates in the range 40-100 000 nGy $h^{-1}$. The device was upgraded with microcontroller PIC16C77, new electronics for amplification and data acquisition system [9].

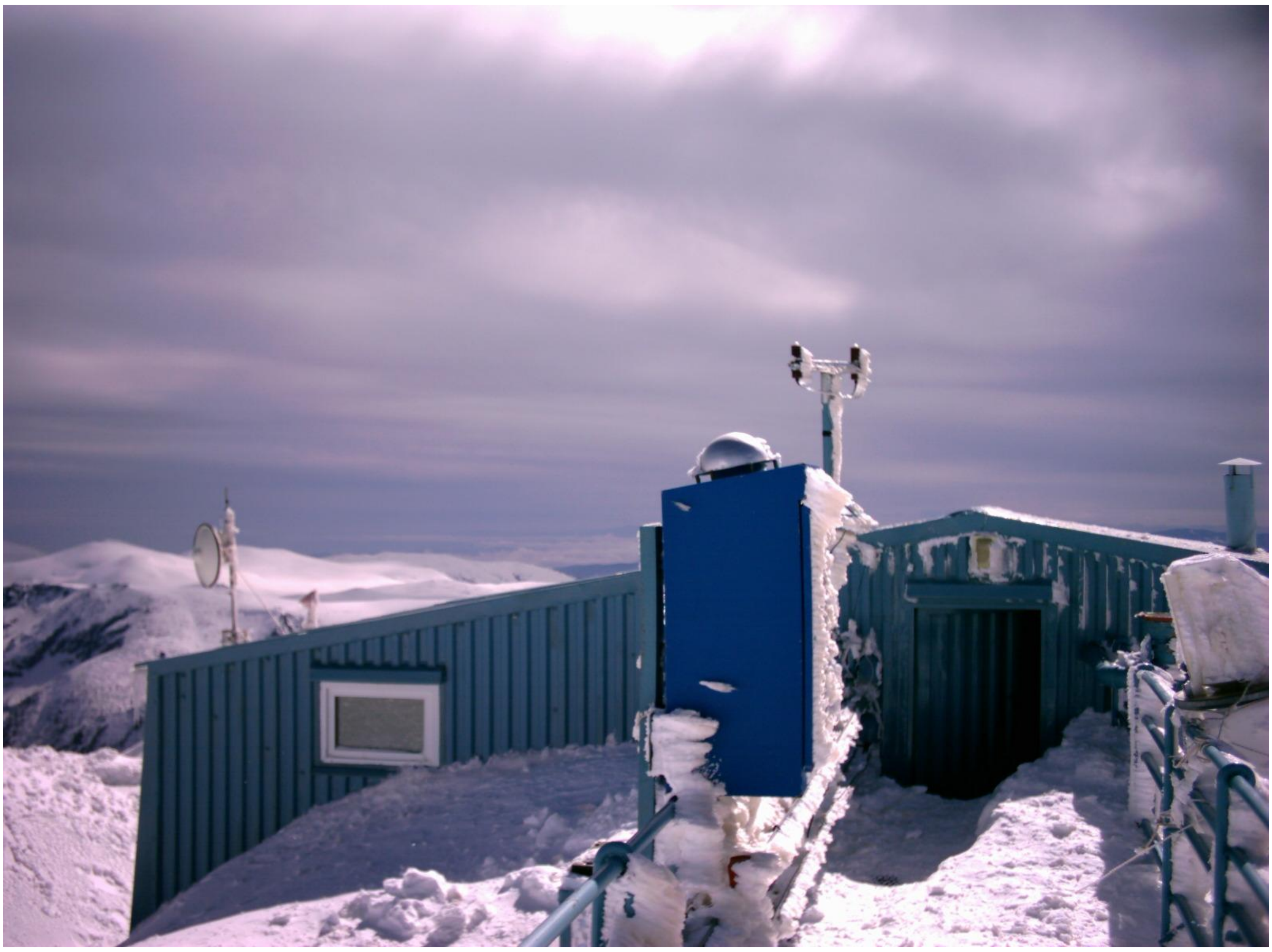

Fig.2 SBN-90 gamma probe based on NaI(Tl) scintillator

The transformation of ionizing energy into an electrical signal is very complicated process and causes a complex temperature dependence of the scintillator's light output [10, 11]. The redistribution of the light intensities is temperature dependent and became significant at temperatures around -20°C [12, 13]. This fact is very important for BEO Moussala, specifically during the winter period (low temperature, high relative humidity, strong winds). Such effect was observed during the 5 years exploitation period of the device, till the end of 2006.

### 2.2 IGS421 gamma probe

The IGS-421 gamma probe is developed by German company TECHNIDATA. It is recently used in the network for continuous gamma background monitoring at BEO Moussala and Institute for Nuclear Research and Nuclear Energy of Bulgarian Academy of Sciences. The probe is based on GM tubes (two low-energy and one high-energy detector). The sensitivity range of the device is 10 nGy $h^{-1}$-10 Gy $h^{-1}$, with accuracy of 15 % respecting to $^{137}Cs$.

The technical characteristics are as follows: The operating temperature is between -40° C and +60° C. This is very important taking into account the negative annual temperature at the top and low temperatures during the winter period (see the previous section). The dimensions of the probe are 80/115mm x 635 mm with weight of ~ 2300 g. The used interface is RS-232, which permits up to 15m direct connection with a PC. The probe is mounted outside of the main bridge of the station (Fig. 3).

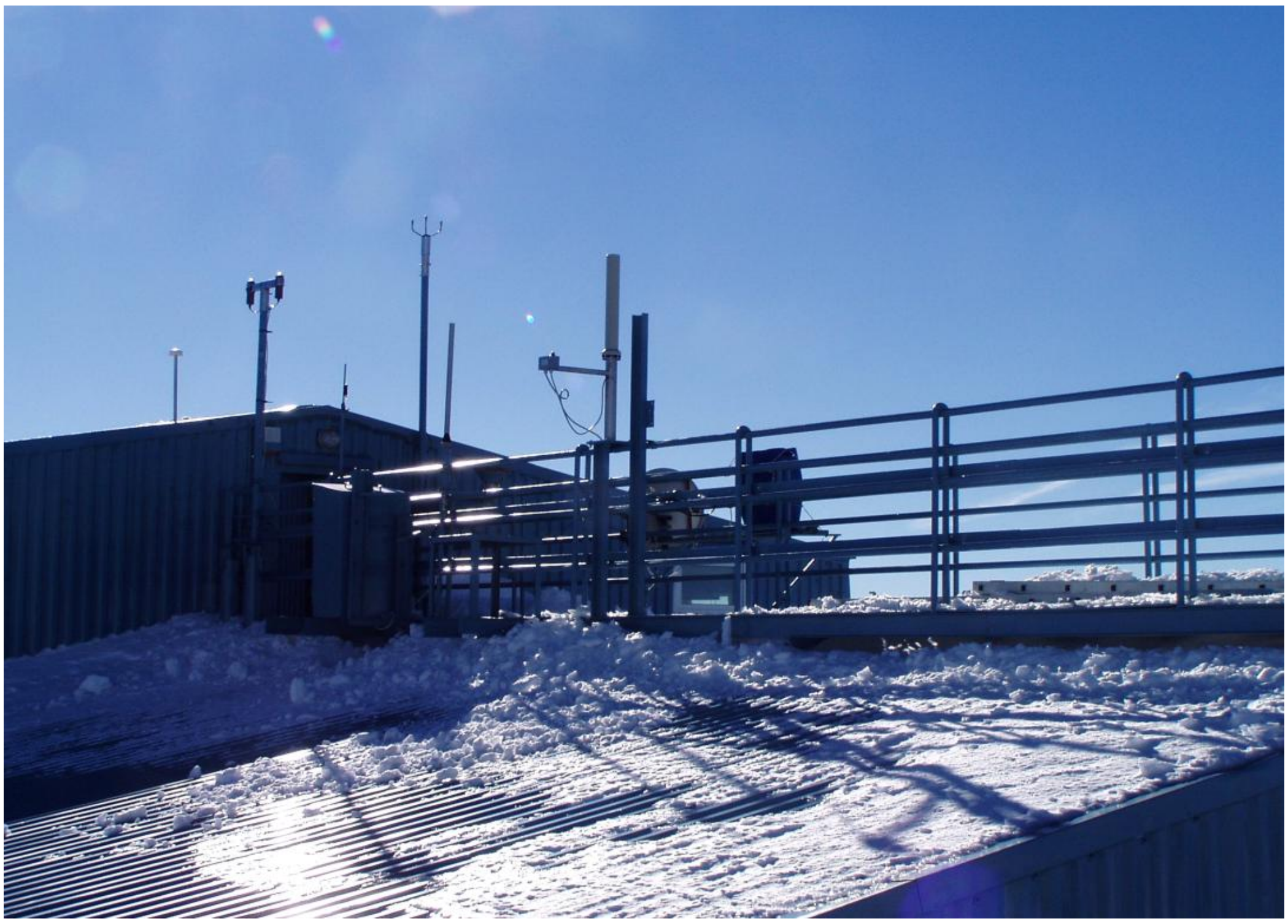

Fig. 3 IGS 421 probe mounted at the bridge of BEO Moussala

As was mentioned above the IGS 421 consists of low-energy and high-energy part. The characteristics of the probe are presented in Table. 1. The probe is operational since September 2006 at BEO Moussala [14].

| Probe | **Range** | **Sensitivity** | **Detector background** |
|---|---|---|---|
| **Low energy** | 10n Gy/h -mGy/h | 1976 counts/min (mGy/h) | 38 counts/min (38nGy/h) |
| **High energy** | 0.1m Gy/h-10 Gy/h | 1.24 counts/min (μGy/h) | |

Table 1. Main characteristics of low energy and high energy GM tubes of IGS 421

**2.3 MDU Liulin**

The mobile dosimetry unit (MDU) Liulin [15] is originally developed for onboard spacecraft studies. It has proved as device for on-Earth radiation measurements, particularly at flight altitudes [16, 17]. The equipment has been acquired for Alomar in the North of Norvege and Jungfraujoch in Switzerland observatories. Quite recently was installed at BEO Moussala station [18, 19]. The spectrometer MDU-Liulin is based on a Si-semiconductor diode. The diode is situated at the head of a MDU unit. It has dimensions of 10x20 $mm^2$, the thickness is 0.3 mm. It is covered by 1 mm of Al and about 0.5mm of copper. There is also an air-gap between Si-diode and Al-cover of about 5 mm thick. The equipment monitors simultaneously the doses and numbers of energy deposition events in Si-diode. The amplitude of the pulses is proportional by a factor of 240 mV/MeV of the energy loss (deposited) in Si. Final adjustment of the energy scale is made through 60 keV photons of $^{241}$Am. The amplitudes are digitized and organized in a 256-channel spectrum. The dose D, in Gy in Si diode is calculated from the spectrum as:

$$D = K * \sum (E_i * A_i)/M_d \tag{1}$$

where $M_d$ is the mass of the detector in [kg], $E_i$ is the energy loss in [J], in the channel i; $A_i$ is the number of events in this channel and K is a coefficient. The experimental time of use of the instrument depends on the power of the battery and on the rate of the memory fills up. As was mentioned above the BEO Moussala MDU unit is concerned and it was operated directly through a computer. Generally the MDU Liulin demonstrated high capabilities for different problems, especially measurements at flight altitudes and registration of solar energetic particles [20-22].

## 3. Measurements of gamma background at BEO Moussala

At present the gamma background at BEO Moussala is measured mainly by IGS-421 gamma probe and Liulin MDU. In this section are presented the data obtained during the first year of exploitation of both devices.

### 3.1 Previous results

The first measurements of gamma background at BEO Moussala are carried out with SAPHYMO gamma probe. The minimal measured dose-rate is 87.50 nSv/h (May 2003) and the maximal is 128.10 nSv/h (July 2004). The average is 118.0 nSv/h with corresponding standard deviation of 7.5 nSv/h. A floating of the measured values, respectively their fluctuations as function of rain flows is observed. The measured dose-rate increases in months with intensive rains. The explanation is that rains collect aerosols from higher level of the atmosphere and transport them to the ground-level.

### 3.2 Recent results from measurements

Presently the gamma background measurements at BEO Moussala are carried out with several different devices. These devices are based on different type of detectors and thus have different accuracy and sensitivity. Since September 2006, the described above devices (IGS-421 and Liulin) are both operational at BEO Moussala. The average of gamma background with the corresponding standard deviations for December 2006 obtained with IGS-421 is presented in Fig. 4. The monthly mean value is 144 nGy/h with standard deviation of 5.19 nGy/h. The maximal measured value is 169 nGy/h, respectively the minimal is 128 nGy/h. An additional analysis of the symmetry of the distribution is carried out. As was expected the observed median value of 145 nGy/h is in practice the same as the mean value. The values are obtained on the basis of 10 minutes integration of the gamma background.

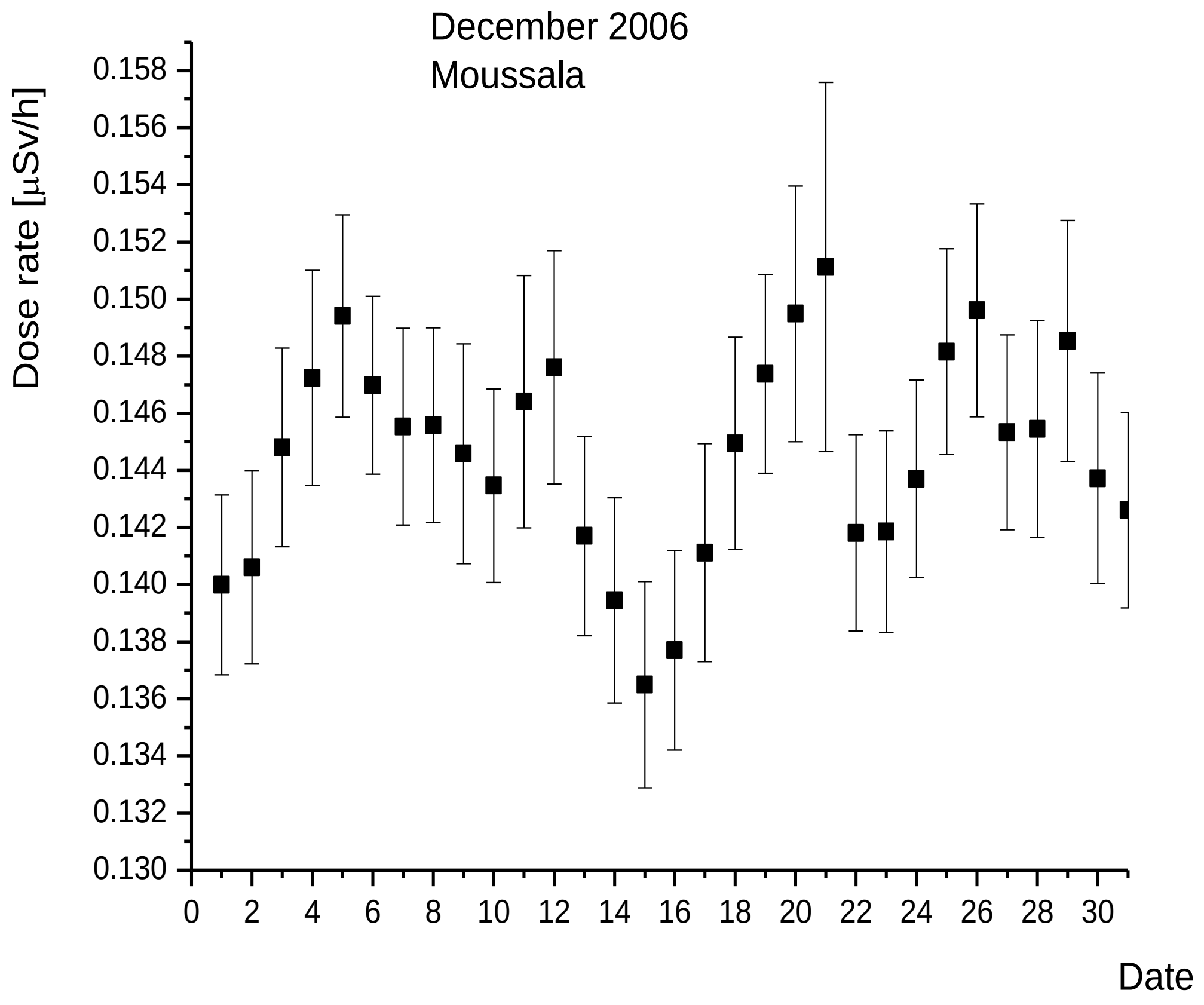

Fig. 4 Average gamma background at BEO Moussala measured with IGS 421 for December 2006

In addition the daily average is presented in Fig. 5. A software improvement was made, aiming the automatic software setting of the probe. As a result in December 2006 (see Fig. 4) the not operational time is reduced to less then 5 % (in a previous months it was in the order of 20%). The not operational regime is due essentially on power supply problems at the top. One of the specifics of BEO Moussala telecommunication system [23] is the direct connection of the probe to a PC. The stored locally data at BEO Moussala are transmitted to BEO-INRNE server via high speed wireless telecommunication system and are presented in a real time on BEO Moussala webpage.

The gamma background at BEO Moussala is estimated also with Liulin spectrometer. The results are presented in Fig.6. The corresponding particle flux is shown in Fig. 7 (see Eq. 1) .

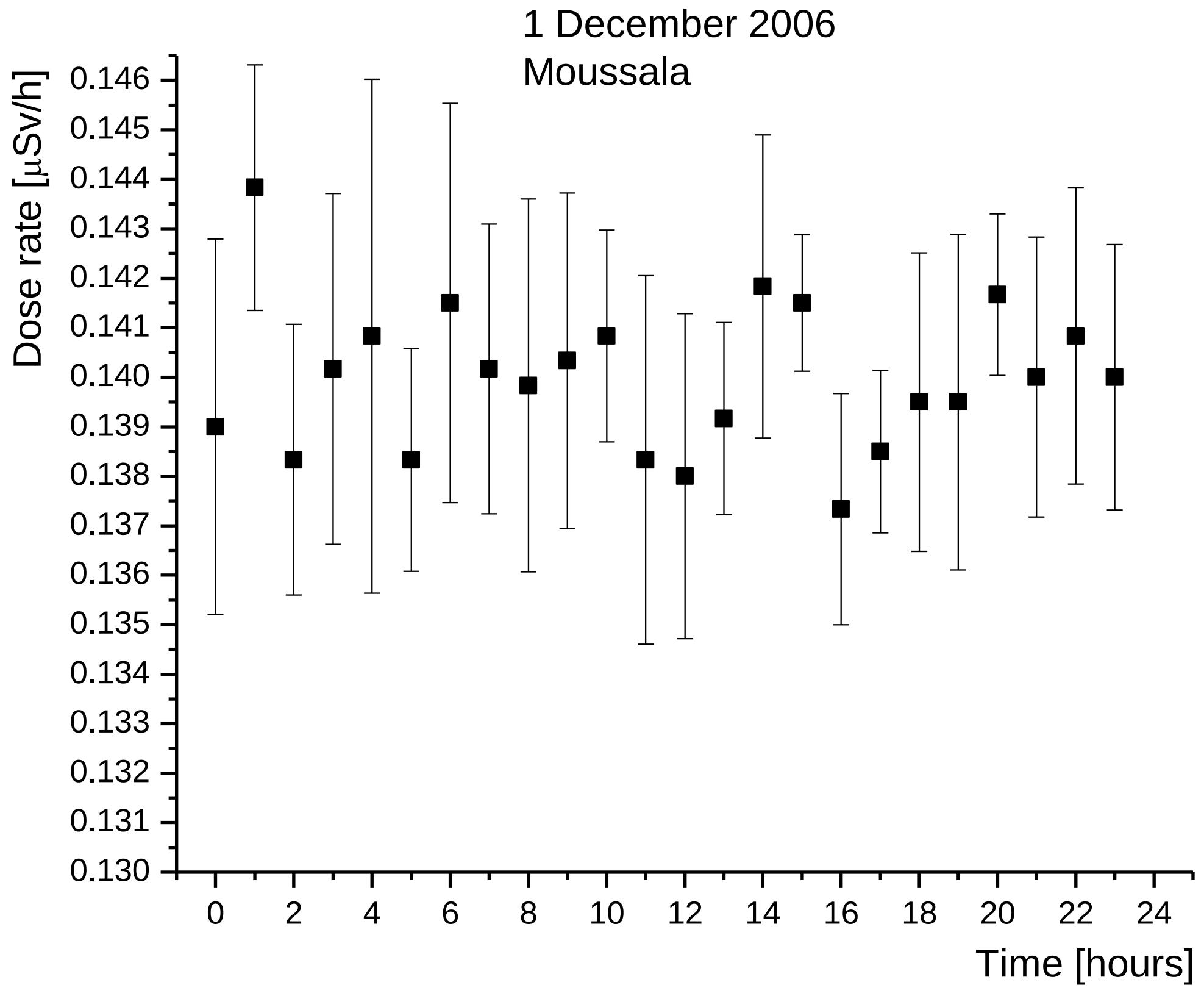

Fig. 5 Gamma background at BEO Moussala measured with IGS 421 for 1st December 2006

A comparison between the measured daily average of gamma background with IGS 421 gamma probe and Liulin for December 2006 is carried out (Fig.8). The monthly average gamma background at BEO Moussala measured with Liulin and IGS 421 gamma probe is shown in Fig. 9. It is necessary to point out the MDU Liulin is located inside the station, while the IGS-421 probe is outside.

In addition similar analysis of the measured data is made. The results are described in Tabl. 2. The measurements carried out with Liulin spectrometer systematically have larger standard deviations. This is due essentially on a smaller detector area of Liulin. The measured with Liulin dose rates are around 20 % greater then measured with IGS 421. An additional comparison with passive TLD detectors measurements is carried out. The dose rate measured with TLDs is 150 nSv/h with deviation of 20 nSv/h (two years data – 3 months integration).

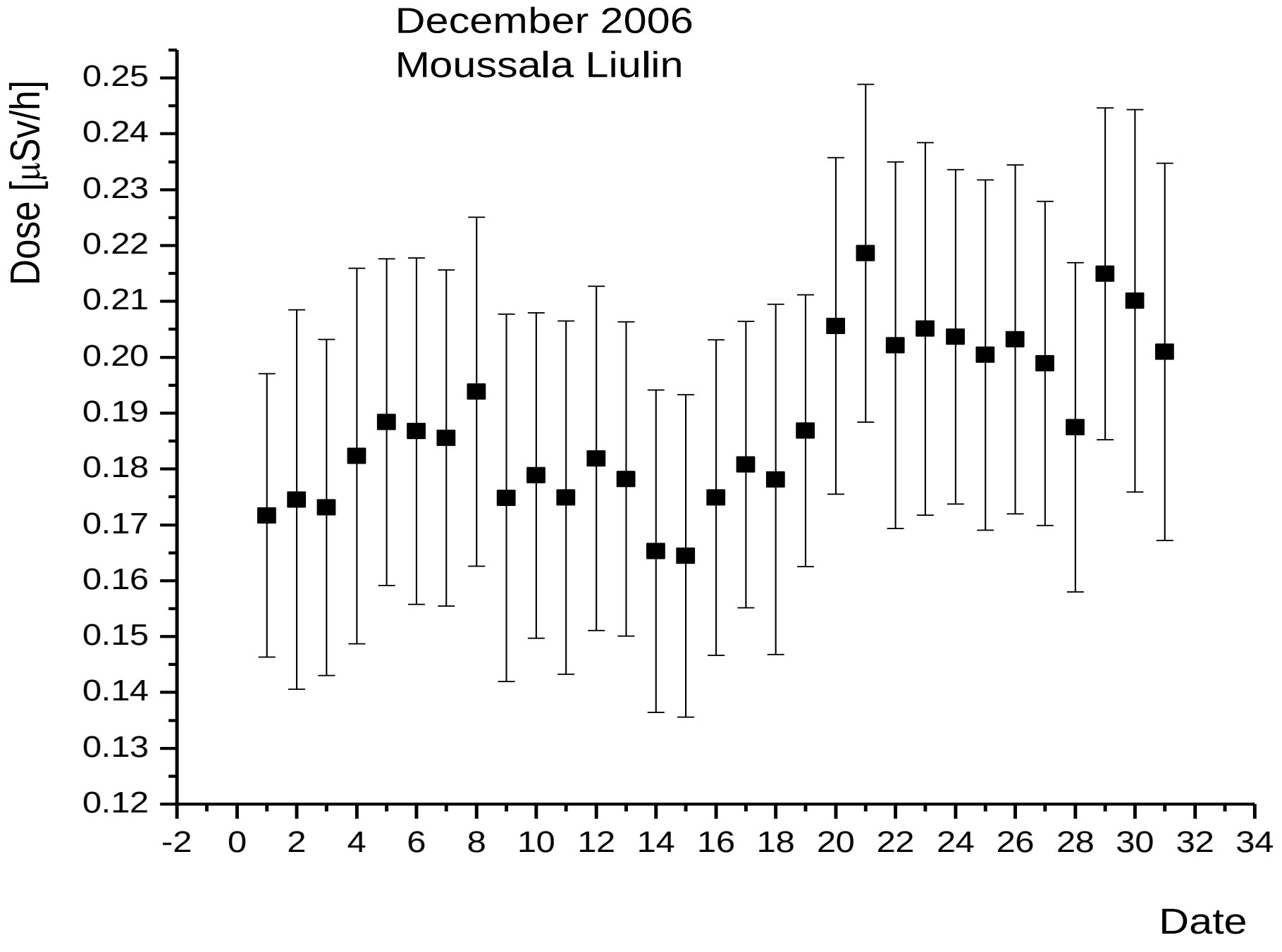

Fig. 6 Average gamma background at BEO Moussala measured with Liulin for December 2006

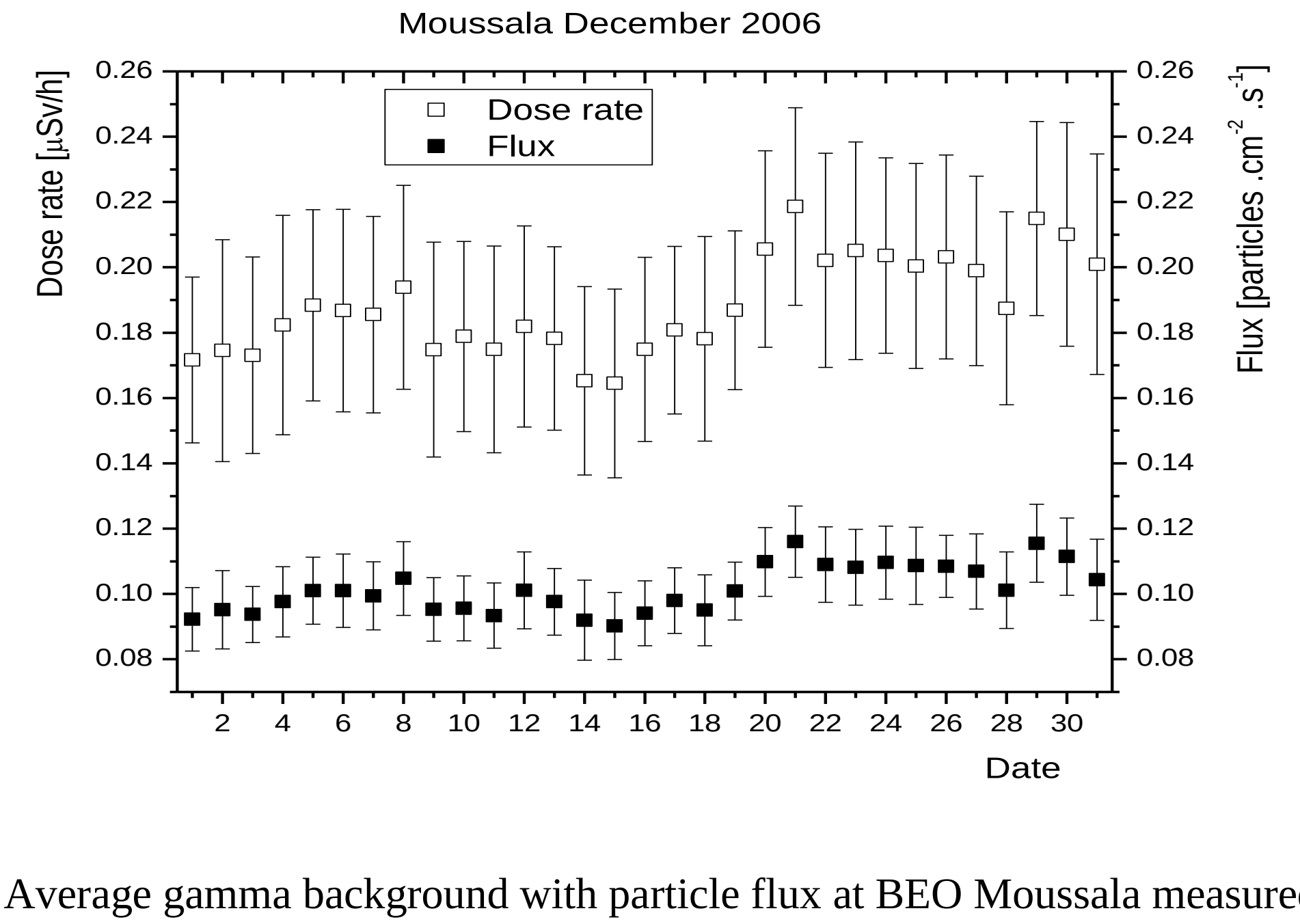

Fig. 7 Average gamma background with particle flux at BEO Moussala measured with Liulin for December 2006

These values are in agreement of the mentioned above data. Generally the described active devices, in addition with TLD detectors show stable and proper work and can be used as a reference point in a Bulgarian national system for permanent gamma monitoring. The IGS-421 measurements are transmitted permanently to EURODEP.

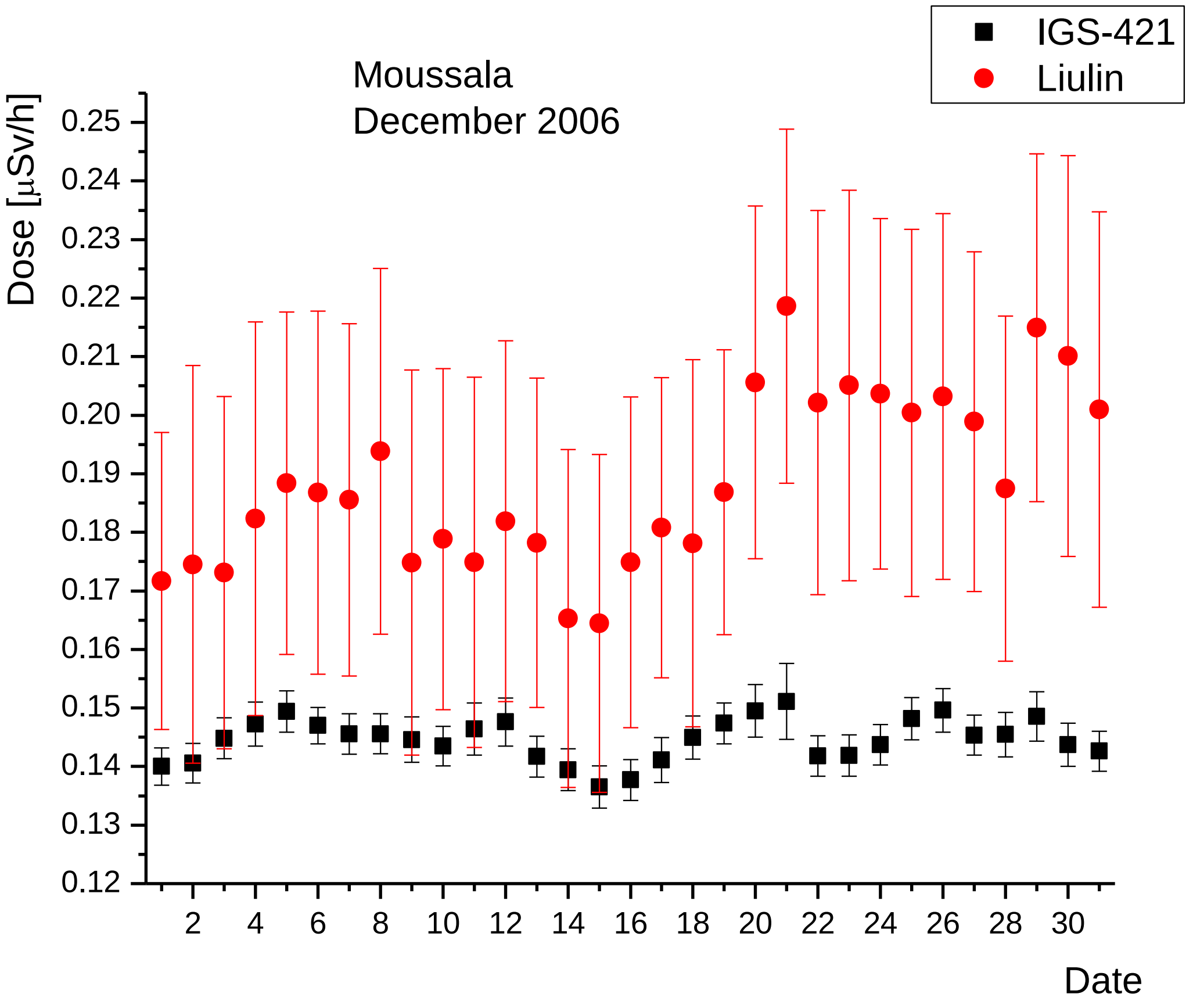

Fig. 8 Average gamma background at BEO Moussala measured with Liulin and IGS 421 gamma probe for December 2006

| Device | Mean [nSv/h] | Standard deviation | Median [nSv/h] | Min [nSv/h] | Max [nSv/h] |
|---|---|---|---|---|---|
| IGS 421 | 144 | 5.19 | 145 | 128 | 169 |
| Liulin | 189 | 33.6 | 186 | 98 | 383 |

Table 2. Mean, median, minimal, and maximal values of the measured gamma background at BEO Moussala with IGS 421 gamma probe and Liulin spectrometer

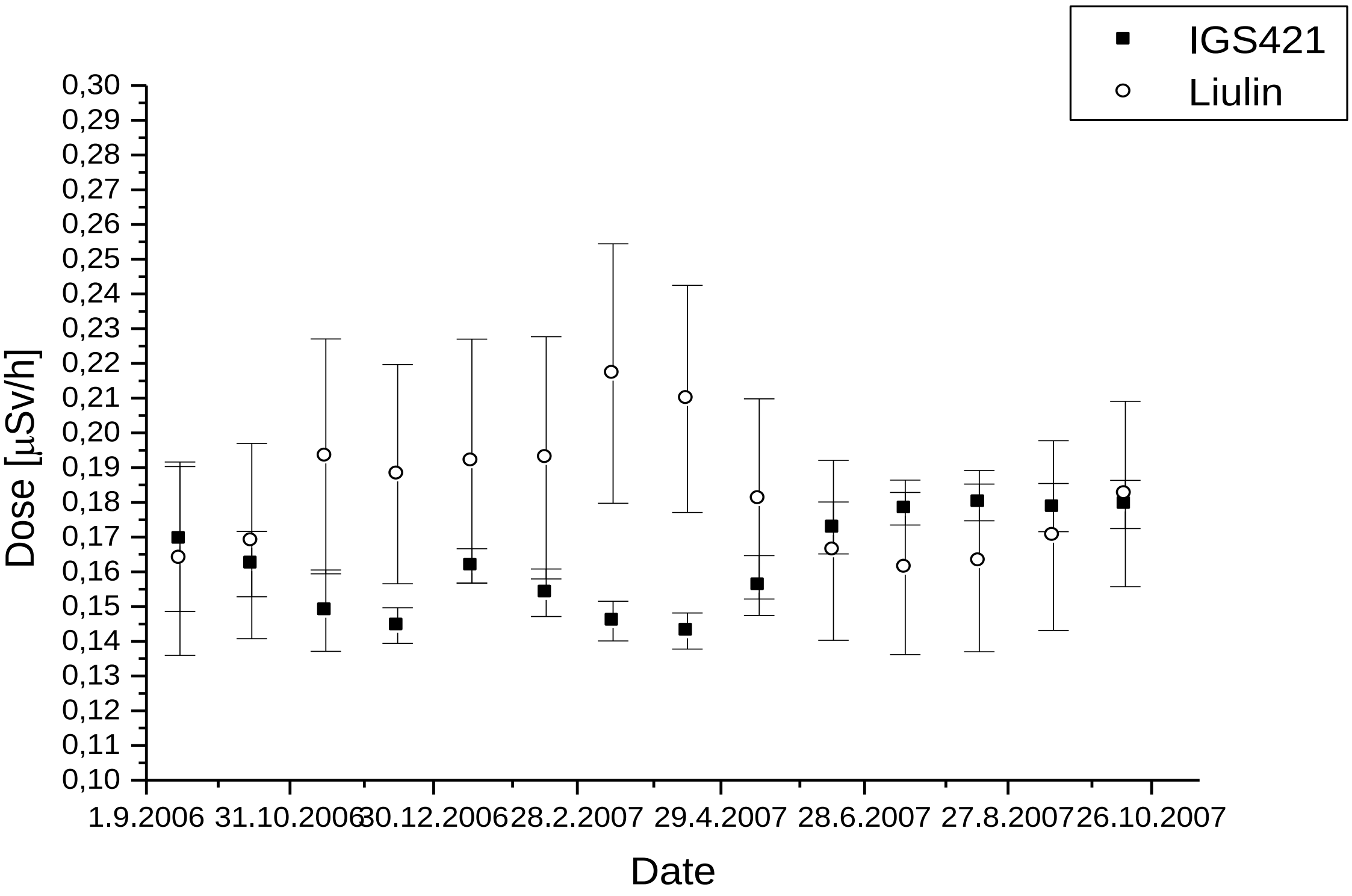

Fig. 9 Monthly average gamma background at BEO Moussala measured with Liulin and IGS 421 gamma probe

## 4. The contribution of cosmic ray to dose rate

The Earth is constantly bombarded by flux of subatomic particles radiation from which outside of solar system-the cosmic ray. Primary cosmic ray particles impinge the Earth atmosphere and release energy via nuclear interaction and ionization losses. Low-energy particles from cosmic ray are absorbed in the atmosphere, while those with energies above GeV generate new particles through interactions with air nuclei, they initiate nuclear-electromagnetic-muon cascades in the atmosphere. The dose from cosmic ray is mainly due to muons, neutrons, and electrons. The dose rate varies with geomagnetic field, respectively geographic position, altitude, and solar cycle (Fig.10).

The estimation of the deposited energy by cosmic ray, respectively dose rate is possible on the basis of a Monte Carlo simulation of the atmospheric cascade. The evolution of atmospheric cascade processes was carried out with CORSIKA 6.52 [24] code with corresponding hadron interaction models FLUKA [25, 26] and QGSJET II. [27].

Cosmic Ray Simulations for KASKADE (CORSIKA) code is one of the most widely used in the last years atmospheric cascade simulation tool in cosmic ray and astroparticle physics. This is a Monte Carlo program for detailed study of cascade evolution in the atmosphere. The code simulates the interactions and decays of nuclei, hadrons, muons, electrons and photons in the atmosphere up to extreme energies.

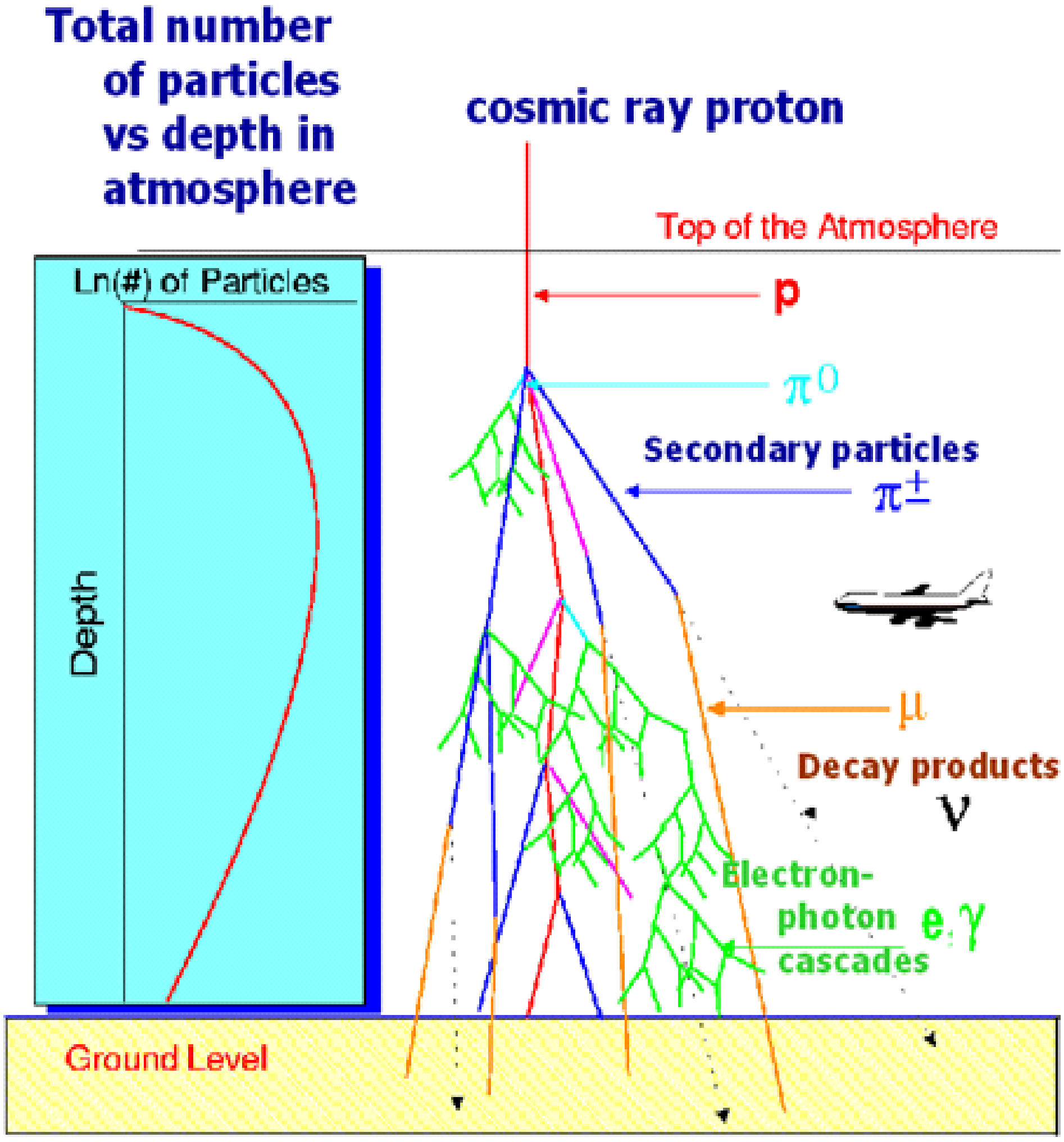

Fig.10 An illustration of cosmic ray induced cascade in the atmosphere with corresponding particle flux

The result of the simulations is detailed information about the type, energy, direction, location and arrival time of the produced secondary particles at given selected observation level. Moreover it is possible to obtain the energy deposit by different shower components and particles at given observation levels. Thus on the basis of the dose rate definition, namely deposited energy in [kg] for given period it is possible to estimate the cosmic ray contribution to dose rate in the atmosphere.

We simulate 50 000 events up to 85 degrees of zenith angle, distributed isotropically, following steep spectrum with index 2.7. The dose rate in [Gy], produced in 1 g of the ambient air at a given atmospheric depth by one particle of the primary cosmic ray with given kinetic energy per nucleon is determined according expression (2).

$$D(h, \lambda_m) = \int_E^{\infty} \int_0^{\pi/2} \int_0^{2\pi} S(E) \frac{\triangle E(h, E)}{\triangle h} \sin(\theta) dE d\theta d\varphi \quad (2)$$

where $\Delta E$ is the deposited energy in a atmospheric layer $\Delta h$, S(E) is the differential cosmic ray spectrum, $\lambda_m$ is geomagnetic latitude, E is the initial energy of the incoming primary nuclei at the top of the atmosphere and h is the latitude above sea level. The geomagnetic latitude $\lambda_m$ governs the rigidity, which is related to integration (integration above E). The dose rate distribution due to cosmic ray as a function of the altitude above sea level is presented in Fig. 11.

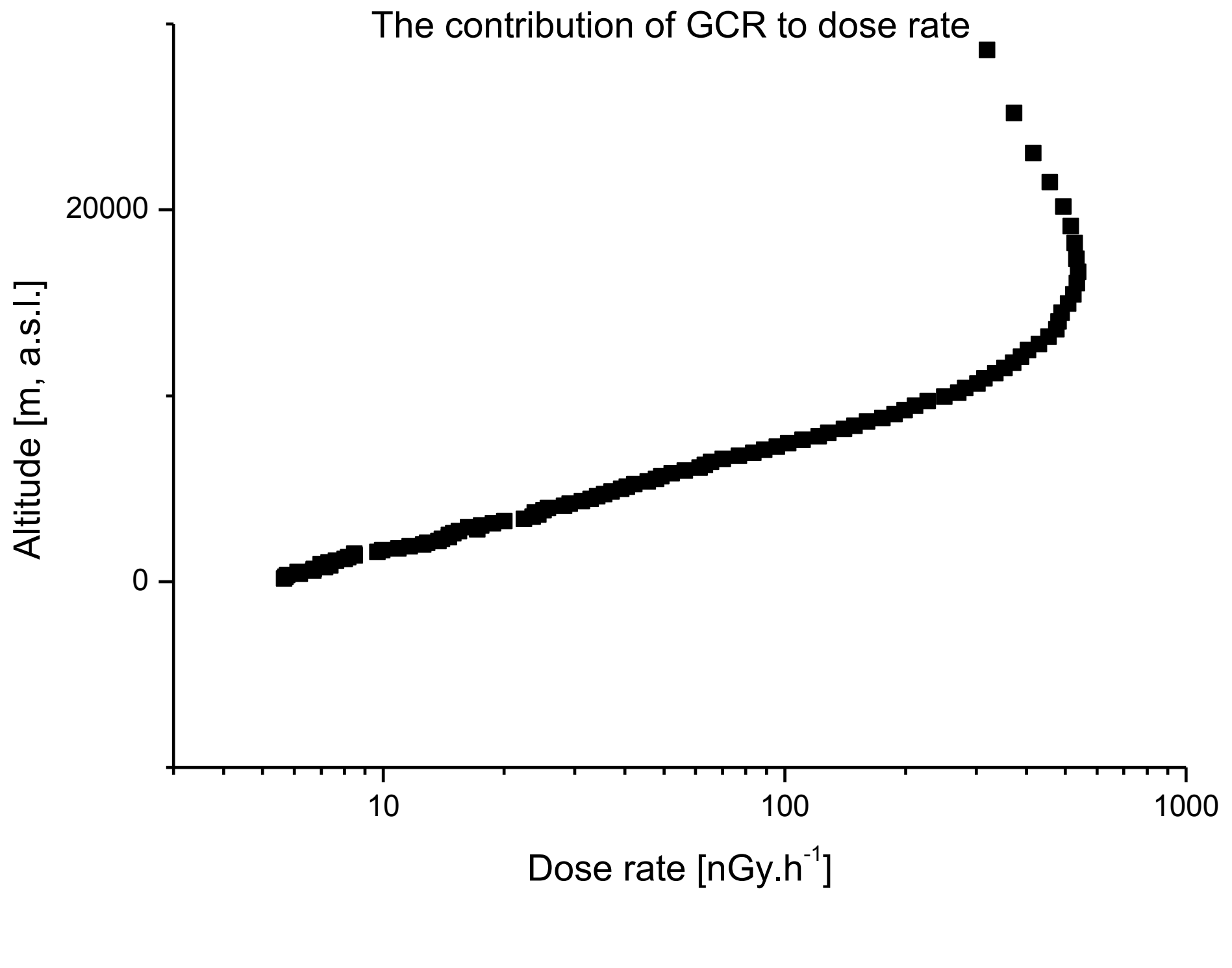

Fig.11 The contribution of GCR to dose as a function of altitude above sea level

The contribution of cosmic ray to dose rate at high mountain altitude is estimated to be of the order of 10-12 % (as example at Moussala is roughly 11 %, see Table 3 and Table. 4). The measured values at Moussala are in full agreement with simulated data and other measurements [27].

| **Depth [g cm$^{-2}$]** | **Altitude [m, a.s.l.]** | **Dose rate from GCR [nGy h$^{-1}$]** |
|---|---|---|
| 813 | 1980 | 12.59 |
| 725 | 2920 | 16.27 |
| 552 | 5000 | 39.2 |
| 365 | 8000 | 128 |
| 272 | 10 000 | 249 |
| 195 | 12 100 | 388 |
| 125 | 15 000 | 509 |
| 75 | 18 200 | 529 |
| 22 | 20 100 | 495 |

## 4. Summary

Presently the gamma background at BEO Moussala is measured with several different devices described above. At the same time on the top is operational automatic meteo-station, which gives permanent information for wind velocity and direction, temperature, pressure and amount of precipitation. The latter is very important, because a slight increase of gamma background was observed in rainy months. As example in Fig. 9 one observes this fact for spring – summer season for 2007, especially for data measure with IGS-421. This fact is not observed with MDU Liulin, because it is located inside the station.

Similar measurements are observed for daily average values. In this work are presented several recent results for gamma background measurements at BEO Moussala obtained with active devices. The work of the new gamma probe IGS 421 is described and the reason to use such type of device for a local network for gamma background monitoring is motivated. The data for gamma background measured with IGS-421 are presented on-line on the web page of BEO Moussala. In addition they are transmitted to EURODEP and national system for permanent monitoring of gamma background in Bulgaria. Additional comparison between several operational devices is carried out. The results are discussed. Moreover the obtained dose rate for BEO Moussala is compared with TLD measurements. Systematically a small increase is observed. The observed discrepancy of the measurements is due mainly to different sensitivity of the devices and not correct calibration of SAPHYMO gamma probe.

On the basis of a full Monte Carlo simulation of atmospheric cascade by cosmic rays is estimated the contribution of galactic cosmic rays to average of dose rate at high mountain latitude. The proposed formalism could be applied for various altitudes and geographical regions.

The contribution of galactic cosmic ray to ambient dose rate is estimated on the basis of a full Monte Carlo simulation of cosmic ray induced nuclear-electromagnetic-muon cascade in the atmosphere of the Earth. Hence the energy deposit by secondary cosmic ray is explicitly obtained. Subsequently the dose rate is obtained using the shape of average primary cosmic ray spectrum. The proposed numerical model could be applied for various geographical regions and also for estimation of dose rate due to solar particle events.

## Acknowledgements

We are grateful to our colleagues from BEO Moussala, software engineer K. Davidkov. This work is dedicated to the memory of Prof. F. Spurny, having significant contribution to improvement radiation measurements on BEO Moussala.